\pdfoutput=1
\documentclass[11pt]{article}

\usepackage[]{acl}

\usepackage{times}
\usepackage{latexsym}
\usepackage{hyperref}
\usepackage{booktabs}
\usepackage{graphicx}
\usepackage[T1]{fontenc}
\usepackage[utf8]{inputenc}

\usepackage{microtype}

\usepackage{inconsolata}

\title{A Survey of using Large Language Models for Generating Infrastructure as Code}

\author{Kalahasti Ganesh Srivatsa\thanks{* Authors contributed equally} \textsuperscript{1},
Sabyasachi Mukhopadhyay\footnotemark[1] \textsuperscript{1,2},
  Ganesh Katrapati\textsuperscript{1}, \\
  \textbf{Manish Shrivastava}\textsuperscript{1}\\
   1. Language Technologies Research Center, KCIS, IIIT Hyderabad, India.\\
   2. Tejas Networks Ltd., Bangalore, India .\\
\texttt{\{kalahasti.ganesh, sabyasachi.m, ganesh.katrapati\}@research.iiit.ac.in} \\
\texttt{sabyasachim@tejasnetworks.com} \\
\texttt{m.shrivastava@iiit.ac.in}
}

\begin{document}
\maketitle
\begin{abstract}
Infrastructure as Code (IaC) is a revolutionary approach which has gained significant prominence in the Industry. IaC manages and provisions IT infrastructure using machine-readable code by enabling automation, consistency across the environments, reproducibility, version control, error reduction and enhancement in scalability. However, IaC orchestration is often a painstaking effort which requires specialised skills as well as a lot of manual effort. Automation of IaC is a necessity in the present conditions of the Industry and in this survey, we study the feasibility of applying Large Language Models (LLM) to address this problem. LLMs are large neural network-based models which have demonstrated significant language processing abilities and shown to be capable of following a range of instructions within a broad scope. Recently, they have also been adapted for code understanding and generation tasks successfully, which makes them a promising choice for the automatic generation of IaC configurations. In this survey, we delve into the details of IaC, usage of IaC in different platforms, their challenges, LLMs in terms of code-generation aspects and the importance of LLMs in IaC along with our own experiments. Finally, we conclude by presenting the challenges in this area and highlighting the scope for future research.
\end{abstract}
 % \\ \newline \Keywords{Infrastructure as Code, Code Generation Models, Large Language Models, Text Generation Models} }

\section{Introduction}

Infrastructure as Code (IaC) \citep{morris2016infrastructure} has gained significance in modern software development as a mechanism for defining and managing IT infrastructure using code-based representations. With features like automation, scalability, consistency and version control \citep{erikFrancisiac}, IaC is revolutionising the provisioning of infrastructure. Developers who are unfamiliar with this technology may find it challenging to create effective IaC templates. Moreover, the manual creation of infrastructure code can be time-consuming, error-prone and challenging to maintain, especially in complex environments.

Large Language Models (LLMs) have emerged as a new paradigm in NLP. Having been trained on a large amount of text for predicting the next word, with the given previous words and sentences by using an in-context learning mechanism, they have shown remarkable performance in downstream NLP tasks such as dialogue modelling, machine translation, question answering, text generation, sentiment analysis and so on.

Additionally, LLMs have demonstrated an ability in tasks related to code generation and validation. For instance, CodeParrot, CodeGen \citep{nijkamp2022codegen}, Llama \citep{touvron2023llama}, Google PaLm and OpenAI's GPT-3.5 \citet{ouyang2022training}, GPT-4 are some of the models which are being studied and utilised for their code-generation capabilities. This leads to the promising possibility that IaC configurations could be generated automatically using LLMs, thereby addressing the problem of a steep learning curve, as well as, enabling users to understand the complexities and adjust parameters accordingly.

\section{Background}
In this section, we give a broad overview of IaC, with an introduction to well-known IaC platforms. This is followed by a brief overview of LLMs, focusing on the applications of LLMs for code generation and then a summary of relevant work regarding the incorporation of LLMs for IaC generation.

\subsection{Infrastructure as Code (IaC)} 
Infrastructure as Code (IaC) is a software engineering and DevOps \cite{bass2015devops} practice that entails managing and provisioning infrastructure through code and automation \cite{punjabi2016user} rather than manual operations. This defines and manages infrastructure elements both in physical and virtual machines, networks, storage and other resources using code scripts or configuration files. IaC tools are categorized \cite{duvall2011continuous} as below two:\par %These scripts or files can be tested, deployed, and versioned like software code, giving infrastructure management automation, consistency, reproducibility and repeatability.
\begin{itemize}
    \item \textbf{Provisioning:} Tools in this category provide infrastructure components for one or more cloud providers. Examples include HashiCorp's Terraform \cite{howard2022terraform} and Pulumi\footnote{\url{https://github.com/pulumi/pulumi}}.
    \item \textbf{Configuration management:} Tools in this category are used for installing and managing software on pre-existing infrastructure. Examples include Ansible\footnote{\url{https://www.ansible.com}}, Chef\footnote{\url{https://www.chef.io/products/chef-infra}} and Puppet\footnote{\url{https://www.puppet.com/}}.
\end{itemize}

\textbf{\textit{Pre-IaC and need for IaC:}} Before IaC, IT had to rely on manual configuration and scripting, which was a tedious process and prone to errors. Additionally, there was a lack of consistency, making it difficult to maintain and troubleshoot. IaC has revolutionized IT management, making it more efficient and reliable than ever before.

\textbf{\textit{A phased approach to the Evolution of IaC:}}

Early IaC methods relied on configurable scripting languages like Shell scripts \cite{greenberg2021unix} but had limitations in version control, modularity, and ease of use. To address these shortcomings, declarative configuration management tools were introduced which also had their limitations in terms of scalability, flexibility, and cloud support. With the advent of cloud computing, the need for dynamic and scalable infrastructure provisioning became apparent. This led to the emergence of infrastructure orchestration tools such as Terraform, Ansible, and Pulumi, enabling on-demand provisioning in cloud environments.

IaC, which employs a descriptive model, follows three key steps: (i) Developers use Domain-Specific Language (DSL) \cite{shambaugh2016rehearsal} to define the configuration state in a file. (ii) This configuration file is then transferred to a server, code repository, or API. (iii) The system configures the infrastructure based on the instructions in the transferred file, ensuring reliable versioning and deployment.

% -----old text-----
% \textbf{\textit{Approaches of IaC:}} There are two main approaches for IaC, Declarative (functional) and Imperative (procedural). The key difference between these two approaches is ``what" vs ``how".\begin{itemize}
%     \item \textbf{Declarative:} Also known as a functional approach, this defines the intended state and then the system carries out the necessary steps to get there without any specific syntax. Examples include Terraform and AWS Cloudformation.

%     \item \textbf{Imperative:} Also known as a procedural approach, this defines the specific commands to the system to be executed in the correct sequence for the intended results. Examples include Ansible and Pulumi.
% \end{itemize}

\textbf{\textit{Approaches of IaC:}} \textbf{Declarative (functional)} approach outlines the intended state, allowing the system to perform necessary steps without specific syntax. Examples include Terraform, AWS CloudFormation. \textbf{Imperative (procedural)} approach provides specific commands in the correct sequence for desired results. Example includes Ansible, Pulumi.

\textbf{\textit{Benefits:}} IaC offers numerous industry benefits \cite{humble2010continuous, cois2014modern}, including automation, consistency, rapid deployment, transparency, version control, scalability, reusability, modularity and immutability. These benefits help in reducing errors, configuration drifts, efficient deployment, tracking the changes, rollbacks, traffic management and reusable code across various environments by ensuring stability without modifying the deployed instances.

% \begin{itemize}
%     \item \textbf{Automation and Consistency:} Automates infrastructure provisioning and management ensuring consistency and reproducible setups, reducing human errors and configuration drift.

%     \item \textbf{Agility and Speed:} Enables rapid and efficient deployment, with easy response to changing business needs and scale resources.

%     \item \textbf{Version control and Transparency:} Enable teams to track changes, collaborate effectively, and roll back to previous versions and transparent for a team to understand the architecture.

%     \item \textbf{Scalability:} Automate the process of scaling the resource for efficient traffic management.

%     \item \textbf{Reusability and Modularity:} Its structure is modular and reusable by adapting across different environments.

%     \item \textbf{Immutable:} Changes are made by creating new instances  without modifying the deployed instances to enhance stability and predictability.
% \end{itemize}

\textbf{\textit{Challenges:}} Though IaC offers the above benefits, it comes with its own set of challenges \cite{siebra2018theory} such as an increase in the complexity due to intricate code for complex configurations, consistency, compatibility, and dependencies across various platforms. It also involves expertise in handling scalability, proper version tracking, tool selection and security aspects.

% Keeping IaC scripts consistent across environments
% \begin{itemize}
%     \item \textbf{Learning curve:} Adoption from manual approaches challenges teams in learning new tools, languages and practices.

%     \item \textbf{Complexity:} Complex configurations can lead to intricate code that requires careful design and testing.

%     \item \textbf{Tool selection:} Selecting an appropriate tool for the task can be a difficult task as there are various tools available.
% \end{itemize}

Some of the popular IaC tools are \textbf{\textit{(i) Terraform}} an open source project, that uses HashiCorp Language (HCL), \textbf{\textit{(ii) Ansible}}, open source project automates provisioning using YAML, \textbf{\textit{(iii) Pulumi}}, a tool with the flexibility for using general programming languages like C, C++, Python etc., \textbf{\textit{(iv) Kubernetes}} etc.

\subsection{Large Language Models (LLM)} 

Language modelling (LM) advances machine intelligence by modelling the generative likelihood of word sequences and predicting future token probabilities. LMs have garnered substantial attention and transitioned through four distinct developmental stages starting from Statistical LM, progressing through Neural LM namely RNN \cite{bengio2000neural, mikolov2010recurrent, mikolov2011extensions}, LSTM \cite{hochreiter1997long} and Transformers \cite{vaswani2017attention}, further advancing into Pre-trained LMs like BERT \cite{devlin2018bert} and culminating in the emergence of LLMs. The Transformer architecture, based on a self-attention mechanism, allows for efficient parallelization and handling of long-range dependencies is a major breakthrough for LLMs. LLMs such as GPT \cite{radford2018improving} and RoBERTa \cite{liu2019roberta} models exhibit high potential in performing NLP downstream tasks.

\subsection{LLMs for Code Generation Task} Code generation, synthesis and summarization tasks have been promising research in recent times with the increased capabilities of LLMs. As per the survey of \citet{xu2022systematic}, there are three ways to pre-train a code generation model. 

\textbf{Decoder-Based LM:} An auto-regressive, left-to-right model also called Causal Language Model (CLM) performs well on code generation and completion tasks. In this, the model predicts the next token based on the previous token. Codex \cite{chen2021evaluating}, a GPT-3 \cite{brown2020language} based 12 billion parameters model which has been pre-trained on 159 GB of code samples from 54 million GitHub repositories, solved 28\% of HumanEval. According to a study by Chen et al., scores have improved with the use of repeated sampling or pass, a concept in which the model is given 100 chances and if it can generate 1 correct sample out of 100 samples, the model has solved the task. CodeParrot\footnote{\url{https://github.com/huggingface/transformers/tree/main/examples/research_projects/codeparrot}}\cite{tunstall2022natural} trained on 25-30B tokens of Google BigQuery data, evaluated on HumanEval where CodeParrot 110M(small) outperformed the CodeParrot 2B(large). CodeGen \cite{nijkamp2022codegen}, proposed Multi-Turn program synthesis model trained on The Pile \cite{gao2020pile}, BigQuery\footnote{ \url{https://cloud.google.com/bigquery}} which has natural language, code, configuration files in its dataset and BigPython datasets. CodeGen-Multi, fine-tuned on Python files and termed as Mono model improved the program synthesis task substantially. Their study says that as there is an increase in the size of the model, there is an increase in the overall performance also. A few more examples of decoder-based models are LLaMA \& LLaMA-2 proposed by \citet{touvron2023llama,touvron2023llama2} are trained on public GitHub data available on Google BigQuery to generate a code based on natural language description. This is evaluated on HumanEval and MBPP \cite{austin2021program} datasets. LLaMA-2 outperformed LLaMA1 and other general models. As per their study further fine-tuning on code-related data would increase the capability of the model. GitHub Copilot is an AI tool developed by GitHub along with OpenAI that takes natural language input and generates, completes and comments the code. Instruct GPT \cite{ouyang2022training} uses Reinforcement Learning with Human Feedback \cite{christiano2017deep, stiennon2020learning} along with models like code-davinci-002, text-davinci-003 that has shown their program synthesis abilities. PaLM \cite{chowdhery2022palm} model takes natural language (NL) prompts and assists in code generation and completion tasks.

\textbf{Encoder-Based LM:} Auto-Encoding model performs well on code detection and classification tasks by utilising information bi-directionally. CodeBERT \cite{feng2020codebert} which is a bimodal pre-trained model trained on natural language (NL) and programming language (PL), achieved state-of-the-art (SOTA) performance on NL-PL downstream tasks by outperforming RoBERTa in a zero-shot setting. CuBERT \cite{kanade2020learning} is another model that outperformed other models with lesser data and fewer epochs.

\textbf{Encoder-Decoder Based LM:} CodeT5 \cite{wang2021codet5} which extends T5 (Text-To-Text-Transfer-Transformer) \cite{raffel2020exploring} works on the objectives of masked span prediction, de-noising sequence reconstruction, and masked identifier prediction with a bimodal dual generation, encourages a better alignment between NL and PL. This outperformed the previous SOTA model PLBART \cite{ahmad2021unified} on all generation tasks including code summarization, text-to-code generation, code-to-code translation, and code refinement.

% This model utilizes both encoder-decoder blocks.

\section{IaC using LLM: Related Works}

There are some related works where IaC has been generated by LLMs. 
\subsection{Ansible-YAML Generation by LLMs} 
\begin{enumerate}
\item \textbf{Ansible-YAML file generation by open-source models:}
The study in \citet{pujar2023automated} explores the use of LLMs transformer-based models to generate Ansible-YAML code from Natural Language prompts, providing an AI assistant for users to increase productivity. IT infrastructure relies on YAML files for defining and configuring crucial elements. It begins by learning from a sizable amount of YAML and Ansible-YAML data, curated from multiple data sources, including GitHub, Google BigQuery, GitLab and Ansible Galaxy \cite{heap2016advanced} and deduplicated using the exact match method. The curated dataset contains closely 1.1M Ansible tasks, YAML playbooks, and nearly 2.2M other generic YAML files. Their pre-trained models WISDOM-ANSIBLE and WISDOM-YAML are trained on CodeGen architecture's checkpoints that contain Ansible-YAML and YAML files to improve the understanding of the syntax and semantics of YAML files. The ANSIBLE-Galaxy dataset is a collection of high-quality files developed and approved by the Ansible community and is utilized to fine-tune the pre-trained models for Ansible-YAML generation tasks. The major contributions of their work are: (i) Providing a formal definition of the problem when applying code generation to Ansible-YAML, (ii) Creating YAML and Ansible-YAML datasets for pre-training and fine-tuning code generation tasks, (iii) Reformalizing the problem of generating Ansible YAML into a code completion task with novel prompts (iv) Proposed two novel metrics designed specifically for Ansible-YAML. WISDOM-ANSIBLE and WISDOM-YAML model's training code is based on Huggingface Transformer library \cite{wolf2020transformers}, with the provided model checkpoints and tokenizers on the YAML data for 9 epochs on 16 A100 GPUs with 80GB of memory, batch size of 32, 5*10\textsuperscript{-5} learning rate and context window of 1024 along with bf16 data type to fasten the training process. The task utilizes a task description consisting of Natural Language prompt X and Ansible-YAML context script C that can generate two kinds of output either a full playbook or a task in the playbook Y. They defined a probabilistic distribution of the Ansible snippet Y given X and C as $p(Y|X,C)$ and the best possible Ansible task snippet is denoted by $\hat{y}=argmax p(Y|X,C)$ 

Thus, the generated Ansible YAML files are evaluated using these 4 metrics \textbf{\textit{Exact Match}}, \textbf{\textit{BLEU}} \cite{papineni2002bleu} and \textbf{\textit{Ansible Aware}} which uses the Ansible YAML syntax knowledge to compare the modules as well as \textbf{\textit{Schema Correct}} which measures the correctness of the result. Results indicate that pre-trained models WISDOM-ANSIBLE and WISDOM-YAML outperform CODEGEN and CODEX(Codex-Davinci-002) on all 4 metrics. Also, their fine-tuned models show an increase in performance with respect to the pre-trained models.

\item \textbf{Ansible-YAML File Generation by Language Models(LM):} A similar study has been done by \citet{kawaguchi2022implementation} where the designed system uses an LM that has been tuned on Ansible playbooks semantics to suggest potential commands based on the status of the current input (i.e., code completion function). The main goal of this work is to prevent network downtime caused by misconfigurations by using an automation tool. This paper proposes an architecture with a client and server paradigm. The client program has an editor for creating Ansible playbooks and a code completion function using LM. In their study, they state that the previous work on completion using LM can only output the candidates of the succeeding command based on the current input command. Suppose when the LM is trained with both YAML configurations, ``tasks:*, name:*, yum:*" and ``tasks:*, name:*, template*", with * indicating variable string, in the previous models when the model is given input as ``tasks:*", it outputs only its immediate command ``name:*" which is a unigram prediction. But in the proposed system, if the model is given an input of ``tasks:*" it outputs the complete Ansible command {``name:*”, ``name:*, yum:*”, ``name:*, template:*”} by recursively using the output candidates of the language model as the input to generate the next output. This candidate list is ordered by the appearance frequency in training data. 

Upon YAML configuration file completion, operator easily sends it to the server application which utilizes Ansible to compile and activate the settings on the target network equipment, while the client program supporting the operator's configuration. The evaluation demonstrates improved accuracy as the number of input commands grows.

% By recursively using the output candidates of the language model as input, the designed system is able to give a complete Ansible command of {``name:*”, ``name:*, yum:*”, ``name:*, template:*”} and the candidate list is ordered by the appearance frequency in training dataset.

% Once the YAML configuration file is complete, the operator can effortlessly push it to the server application. The server program uses Ansible to compile the received YAML file and activate the setting on the target network equipment, while the client program supports the operator's configuration. The evaluation shows that the system can propose more correct candidates as the number of input commands increases.

% Upon YAML configuration file completion, operator easily sends it to the server application which utilizes Ansible to compile and activate the settings on the target network equipment, while the client program supports the operator's configuration. The evaluation demonstrates improved accuracy as the number of input commands grows.

% Upon YAML configuration completion, the operator seamlessly sends it to the server application. The server, utilizing Ansible, compiles and enacts these settings on the target network equipment, supported by the client program. The evaluation demonstrates improved accuracy as the number of input commands grows.
\end{enumerate}

\subsection{LLMs used in DevsecOps}
\begin{enumerate}
\item \textbf{Static Code Analysis of IaC:} The primary goal of \citet{petrovic2023chatgpt} is to leverage ChatGPT for static code analysis in order to target different IaC standards, with a particular emphasis on Terraform and Ansible in the context of DevSecOps.
\vspace{-1mm}
As a first step, the user must choose and upload the desired archive that contains the IaC scripts in charge of deploying the underlying infrastructure. Each individual IaC file is read and converted as a string to form the pre-defined question for the ChatGPT in the form of the question “Find security flaws in ${filetype}$ script:${contents}$”.
Here, the first parameter serves as a placeholder for the IaC-related file type and the second one contains the script's actual content. Following the construction of the inquiry, a ChatGPT request is submitted via Python API and then a summary of the received responses for each of the IaC files is included in an HTML table that serves as the final output of the solution. Each entry represents a different file from the archive, along with ChatGPT's summary with probable defects and suggestions for resolving it.

\item \textbf{Run-time Analysis of Server Logs:} The study \citet{petrovic2023machine} primarily focuses on leveraging a machine learning approach using server log analysis to detect suspicious activities in runtime security through traffic-related data. Using ChatGPT, this novel technique utilizes context i.e labelled data and queries, in conjunction with log entries to assess the suspicious activities. An essential aspect is the volume of data required for training new prediction models when log structures change. This study explores the innovative application of ChatGPT and LLM for log analysis, aiming to reduce the need for extensive training data by adapting a pre-trained model to yield satisfactory results. The user input comprises question-context pairs, where the context consists of group of labelled log entries, serving as an input to ChatGPT for pattern extraction. With this input, ChatGPT labels network traffic records and explains the meaning of log records, including its underlying protocol, data exchange, communication flow and network traffic details. Finally, the user receives notifications about any detected suspicious traffic or activities.
\end{enumerate}

% \item \textbf{Run-time Analysis of Server Logs:} A study \cite{petrovic2023machine} proposes a machine learning-based approach using server log analysis to identify suspicious activities in run-time security using data connected to traffic. A novel technique using Python's ChatGPT utilizes context (labelled data and questions) and log entries to assess the indication of suspicious activities.

% The research focuses on identifying suspicious activity at run-time using machine learning predictive models against log records. The quantity of information needed to train a new prediction model each time when the log structure changes is one important consideration in this area. The research investigates and addresses the usage of ChatGPT's novel LLM for log analysis in the form of question-answer conversations with the objective of using lesser training data and adapting a pre-trained model for providing satisfactory results.

% The user's input represents a pair that is given as question and context where context means a group of labelled log entries which is sent to ChatGPT as sample data for pattern extraction. Based on the input, ChatGPT labels the network traffic record and also explains the meaning of the given log record, along with the underlying protocol, data exchange, and extra information pertaining to communication flow and network traffic. Finally, the user will be informed if there is any suspicious traffic or activity.

\subsection{IaC Generation through ChatGPT Queries}
\begin{enumerate}
% \item \textbf{ChatGPT for DevOps:}
% In this blog, \cite{CaoimheHarveyDevOpsChatGPT} the author explained the use of ChatGPT for DevOps for prompting the ChatGPT tool to produce small scripts in Python and Bash for which the results were accurate with code explanation. With the results generated, the author queried Chatgpt for some task-specific Terraform configuration, the configuration generated by the OpenAI Playground was identical to the manual configuration and comparatively better than the ChatGPT's result.

% The author continued with a few more prompts to ChatGPT with respect to DevOps concepts. Though the results were not so accurate and in fact similar to the Google search results, the usage of ChatGPT or OpenAI playground was helpful in understanding and getting a way out of some of the not only in CI/CD but even for debugging the code. 

\item \textbf{ChatGPT for DevOps:} The blog \citet{CaoimheHarveyDevOpsChatGPT} demonstrates use of ChatGPT for DevOps by prompting the tool to show its ability to generate accurate Python and Bash scripts. With the results generated, the author queried the task-specific Terraform configuration using Chatgpt, finding that OpenAI Playground's configuration was identical to manual configurations and better than ChatGPT's results. However, in queries about broader DevOps concepts, results were less accurate, resembling Google search results. Nevertheless, utilizing ChatGPT or OpenAI Playground proved helpful in understanding and getting a way out for some of the CI/CD and code debugging tasks.

% \item \textbf{SSO in Kubernates Configuration Generation:} In the blog \citet{BenoitChaGPTvsDev}, ChatGPT was used to generate SSO in Kubernates configuration in Azure, also kube-apiserver manifest details were queried for configuration for SSO using Azure Active Directory, it was seen that ChatGPT generates an example of a kube-API server YAML file and explains each parameter related to SSO. ChatGPT was queried to give an example of a kube-config file to use with the kube-apiserver, which it was able to generate. Next, it was asked to modify the kube-config by using oidc-login plugin of kubectl, which was able to give a proper solution.

% However, it is seen that when used as a solution to connect the oidc plugin with HTTPS to Azure, it gives an additional parameter ``–https" which doesn’t exist. When challenged, ChatGPT acknowledges this mistake and points to the direction of a reverse proxy, which is different from the actual desired parameters. 

\item \textbf{SSO in Kubernates Configuration Generation:} In the blog \citet{BenoitChaGPTvsDev}, ChatGPT was utilized to generate Kubernetes configurations for Azure, specifically related to Single Sign-On (SSO) configurations using Azure Active Directory. ChatGPT successfully generated a sample kube-API server YAML file and detailed its parameters related to SSO and also a sample kube-config file for use with the kube-apiserver. Although it provided a valid solution using the oidc-login plugin of kubectl, an error surfaced when connecting the oidc plugin with HTTPS to Azure, suggesting an extra, non-existent parameter "-https." When challenged, ChatGPT acknowledged the mistake and advised using a reverse proxy, that is different from the desired configuration parameters.
\end{enumerate}

\subsection{IaC Generation Tools with LLM}
\begin{enumerate}
% \item \textbf{Infracopilot \cite{claudioinfracopilot}}

% A cutting-edge IaC editor called InfraCopilot \cite{claudioinfracopilot} is revolutionizing how cloud infrastructure is developed and managed. It uses Klotho\footnote{\url{https://klo.dev/announcing-infracopilot/}}, an open-source engine, that provides unparalleled intelligence, flexibility, and agility to create and change cloud architectures. 

% Apart from the Klotho engine, The InfraCopilot service has several other components like API/Orchestrator, Intent Parser, Visualization Engine, and Discord Bot. To communicate requests to the service, the user engages with the Discord Bot. LLMs send the extracted user intent to the intent corrector, which confirms, corrects and converts it into a JSON format. The updated user intent is then expanded into a verified architecture by the Klotho Engine. All of the low-level components, such as VPCs, subnets, security groups and IAM policies are a part of the multi-level architecture that the Klotho engine produces.

% In contrast to other LLM design generators, InfraCopilot only uses LLM to comprehend user intent and Klotho Engine architecture to provide consistent, clear and reliable infrastructure creation, modifications and upgrades.

\item \textbf{Infracopilot} \cite{claudioinfracopilot} is a cutting-edge IaC editor that uses LLM to understand user intent and Klotho \footnote{\url{https://klo.dev/announcing-infracopilot/}}, an open-source engine, to revolutionize cloud infrastructure development and management by providing unparalleled intelligence, flexibility, agility, consistency, clarity and reliability. It includes components like API/Orchestrator, Intent Parser, Visualization Engine, and Discord Bot to communicate requests to the service. LLMs extract user intent, sent to intent corrector, that confirms, corrects, convert it into JSON, and update intents. Klotho Engine generates and verifies architecture, including low-level components like VPCs, subnets, security groups, and IAM policies.

\item \textbf{K8sGPT} \cite{jasbirK8SGPT} is a tool that scans and diagnoses Kubernetes clusters using SRE-encoded analyzers like PodAnalyzer, pvcAnalyzer, rsAnalyzer, serviceAnalyzer, eventAnalyzer and ingressAnalyzer providing efficient troubleshooting. It features default and customizable analyzers like  hpaAnalyzer and pdbAnalyzer which are activated for specific needs. Filters control resource analysis, enhance analysis capabilities with commands like "k8sgpt filters list" displaying available filters and "k8sgpt filters add/remove" for adding or removing multiple filters. "k8sgpt integration activate/deactivate" activates or deactivates the integrations with tools like Trivy.

% \item \textbf{Pulumi AI\footnote{\url{https://www.pulumi.com/ai/}}} introduced an AI Assistant to speed up the process of finding, comprehending and using the cloud infrastructure API for developing cloud infrastructure using LLMs and GPT. In order to enable intelligent resource identification and interaction within the Pulumi cloud, Pulumi Insights \cite{pulumiinsights} provides IaC intelligence using generative AI and LLMs for enterprise, search and analytics spanning infrastructure and cloud.

% Additionally, Pulumi Insights offers teams crucial components for managing cloud footprints using AI and LLMs. With a Pulumi resource supergraph displaying metadata and linkages across cloud infrastructure, it facilitates learning, finding and constructing cloud architecture. Using their own data warehouse and BI tools, users can visualize Pulumi resource data for cost, compliance and operational use cases.

\item \textbf{Pulumi AI\footnote{\url{https://www.pulumi.com/ai/}}} introduced an AI Assistant to accelerate cloud infrastructure development by leveraging LLMs and GPT, aiding intelligent resource identification and interaction within Pulumi cloud. Pulumi Insights \cite{pulumiinsights} deploys generative AI and LLMs for infrastructure enterprise, analytics and offering key elements to manage cloud footprints. The Pulumi resource supergraph provides metadata and links across cloud infrastructure, assisting in cloud architecture design, allowing users to visualise data for cost, compliance, and operations using preferred BI tools.

\end{enumerate}

\section{IaC Code Generation Process}
This section outlines a template for the generation of Infrastructure as code through LLMs using Terraform as an example and some basic results.
\subsection{Pre-Training} 
Pre-trained models \cite{qiu2020pre} serve as foundational components for diverse downstream tasks, trained on large benchmark data to facilitate easy fine-tuning in various applications and tasks, enabling comprehensive knowledge capture and semantic representation for tasks like text generation etc.

% Pre-trained models \cite{qiu2020pre} in the NLP serves as the fundamental building blocks for a wide range of downstream tasks, each involving diverse data modalities. This technique trains the model using large benchmark data and tasks for easy fine-tuning in various applications that make it easier to solve new tasks. This enables well-trained LMs to capture and learn the rich knowledge and semantic representation by utilizing unlimited training data from unlabeled text corpus for downstream tasks like text generation etc. 

%In this, the model is trained on vast amounts of natural language by fine-tuning the models on specific programming tasks that generate code based on natural language descriptions of a desired program.

% \subsection{In-Context Learning, Indexing, and Prompt Tuning}
\subsection{In-Context Learning}
% \begin{itemize}
% \item \textbf{In-Context Learning:}
In-context learning, also often called ``zero-shot'' or ``few-shot'' learning, utilizes the pre-training data to provide context and task examples, enabling models to infer expected behaviour and produce appropriate responses. This concept helps in rapid experimentation without fine-tuning model settings in unlabelled data situations.

For generating \textbf{\textit{``text to terraform configuration''}} with a few-shot setting in a model, it utilizes the text prompt for specific configuration and context files with sample terraform configuration as an input and the model keeps repeating until all blocks of terraform are generated. In our example, we used a  model with a few-shot setting in LLM like GPT 3.5-Turbo for generating ``text to terraform configuration'', and used context files with sample terraform, provider and resource blocks along with a text prompt given for a specific configuration and the model generates the configuration for the block, repeating until all blocks are generated.

This offers numerous advantages, particularly in situations that lack labelled data or require UI/API interaction, allowing rapid experimentation without fine-tuning model settings.

\subsection{Data Collection and Fine-Tuning}
\begin{itemize}

%-----for old content regarding IFT uncomment below------
% \item \textbf{Instruction Fine-Tuning:} LLMs prompted for generating IaCs might exhibit unanticipated behaviours, such as fabricating information or generating harmful content, due to the current objective of ``Generating IaC with an IaC prompt", being out of alignment with the pre-trained language modelling objective of ``Predicting the next token" \par

% Therefore, it is essential to be fine-tuned for these tasks. The process of fine-tuning involves identifying the task for designing the ideal architecture, loss function, and data selection. Thus our work, Generating ``Infrastructure as Code'' configuration files from English text will be a powerful tool for engineers to refine configurations after the review.

\item \textbf{Instruction Fine-Tuning:} It is imperative for LLMs when generating IaCs with an IaC prompt due to its misalignment with the pre-trained language modelling objective of predicting the next token, as it might exhibit unanticipated behaviours like generating harmful or fabricating content. Hence using fine-tuning approach is essential for these tasks that involves defining the task, architecture, loss function, and data selection. Thus our work, generating ``Infrastructure as Code'' configuration files from English text empowers engineers to refine configurations post-review.

%----not neccessary----
% As a result, these need to be fine-tuned and for fine-tuning initially the fine-tuning task need to be indentified, which helps in designing the ideal architecture, loss function and selecting relevant data for fine-tuning. For the work focuses on generating "Infrastructure as Code" configuration files from English text will be a powerful tool for engineers, to generate and refine the configuration after the review. \par 

%-----for old content regarding IFT uncomment below------
% As a next step collect the data specific to the task. To generate IaC for Terraform, run a query on Google BigQuery GitHub dataset with a .tf extension, which contains over 164MB of Terraform data i.e. 23839 files out of 1.5TB of the corpus. Remove duplicate files, perform pre-processing and divide into training and validation sets in a 75/25 ratio. For our experiment with Code-parrot 19071 files were used for training and 4768 files for validation.\par

To initiate fine-tuning, data pertinent to the task must be collected. For Terraform IaC, Google BigQuery's GitHub dataset with .tf extensions serves as a substantial source, comprising 164MB of Terraform data across 23839 files from a 1.5TB corpus. After eliminating duplicates, we pre-processed, and split the data into 75\% training and 25\% validation sets. In our Code-parrot's experiment we utilized 19071 files for training and 4768 for validation.

%-----for old content regarding IFT uncomment below------
% The pre-trained model that learnt linguistic features from vast code and text corpus must then be loaded and can be adapted for fine-tuning. The adapted model for fine-tuning entails a continuous training process on task-specific data in our case the terraform prompts and configuration examples. We gently keep modifying or tuning the pre-trained parameters of the model to better suit the current task and use gradient descent for updating parameters on task-specific data loss. Performance can be improved by adjusting hyper-parameters such as training epochs, batch size, learning rate, and weight decay. The improved model is kept for analysis. For our example with Code-parrot, the fine-tuning was done for 20000 epochs with the datasets mentioned in the above section. 

The pre-trained model, enriched with linguistic knowledge from vast code and text data, is then adapted for fine-tuning. The adapted model involves continuous training on task-specific data that contain terraform prompts and configuration examples in our work. We adjust the model's pre-trained parameters to better suit the task, using gradient descent to update parameters based on task-specific data loss. Performance enhancements are achieved by adjusting hyper-parameters like training epochs, batch size, learning rate, and weight decay. The refined model is preserved for analysis. In the Code-parrot example, fine-tuning spanned 20000 epochs by utilizing the above datasets.

\end{itemize}

\subsection{Evaluation}
In order to benchmark generative models for code, samples are typically compared to a reference solution for Functional correctness \cite{chen2021evaluating}
%ensures that functionally equivalent configurations that look different are accounted for.\par
A similar approach is used for IaC evaluation too. 
\begin{itemize}
\item \textbf{Functional correctness by exact match:} This metric compares the function of a generated configuration file to a reference solution by ensuring the same functionality across different setups.
A Terraform uses LLM-generated configuration file to develop and create a JSON execution plan if the generated configuration file compiles and matches with the reference solution plan and considered as a success. This implies that even the slightest error in the configuration file generated is considered as a failure. The reference dataset contain tasks that have a natural language description and a desired configuration. Terraform 1.4.6 was used for this activity.

% \item \textbf{Functional correctness by exact match:} This metric compares the function of a generated configuration file to a reference solution by ensuring the same functionality is considered across different setups.
% The LLM-generated configuration file will be used by terraform to develop an JSON execution plan. A JSON file containing the plan is created if the generated configuration file compiles, and if it matches the reference solution plan, the generated configuration file is deemed successful. This implies that even the slightest error in the configuration file generated will be considered a failure. The reference dataset contain tasks that have a natural language description and a desired configuration. Terraform 1.4.6 was used for this activity.
\end{itemize}

% The LLM-generated configuration file will be used by the IaC tool to develop a JSON execution plan. The process of evaluation is two-step as below:
% \begin{enumerate}
% \item The generated configuration file should compile this metric by determining whether the IaC tool successfully built a plan from the configuration file and whether the right sort of resource was created.
% \item The compiled configuration file should match the reference solution plan for the task.
% \end{enumerate}

% The evaluation dataset covers all features of Terraform configuration files including jobs from GCP, AWS, and Azure, as well as ranging from virtual machines to VPNs. The task complexity varies widely in order to push the LLMs to their maximum capability.

% All facets of configuration file generation, including the main cloud providers(AWS, Azure, and GCP) should be covered by these tasks. All different Infrastructure configurations covered by the different IaC tools for example Virtual Private Clouds (VPCs), virtual machines, firewalls, Virtual Private Networks (VPNs), security groups, Kubernetes PODs etc should be examined. The complexity of The task varies widely in order to push the LLMs to their maximum capability\par

A task is is a JSON-formatted text file that describes one or more containers that forms our application.

The evaluation dataset covers all features of Terraform configuration files including jobs from GCP, AWS, Azure and virtual machines with task complexity varying to maximize LLMs capabilities. To compare two plans, JSON file details need anonymization, and multiple samples are generated per task to determine the average success rate due to the stochastic nature of code/config generation.

% The identities, names, descriptions, and timestamps should be deleted from the JSON files before the two plans are compared. As code/config generation is stochastic, for each task multiple samples should be generated to determine the average success rate.

% \begin{itemize}
% \item \textbf{Functional correctness by exact match:} Functional correctness metric compares the function of a generated configuration file to a reference solution. It also ensures that different setups but the same functionality are taken into account.
% The created configuration file will be used by Terraform to try and develop an execution plan. A JSON file containing the plan is created if the generated configuration file compiles, and if it matches the reference solution plan for that task, the generated configuration file is deemed successful. This implies that even the slightest error in the configuration file generated will be considered a failure. Terraform 1.4.6 was used for this activity.

%\item \textbf{ROUGE:} Variants of ROUGE \cite{chin2004rouge} like ROUGE-1, ROUGE-2, ROUGE-L, recall, precision, and F1 scores can be used to measure the match between configuration file execution plans.

%\item \textbf{BLEU:} By comparing the execution plan of the generated configuration file to the required configuration, BLEU \cite{papineni2002bleu} determines the average precision over a range of n-gram sizes. 
% \end{itemize}

\subsection{Experiments and Results Analysis}

As a part of the survey, we performed experiments with 4 GPUs of Nvidia GeForce RTX 2080 Ti (11GB) and reported our scores on the models CodeParrot small (110M) and GPT-3.5-turbo models with a single sample and multiple sample configuration generation for 49 different tasks for AWS service providers. These tasks are a collection of various configuration areas on the AWS cloud through Terraform. The Table~\ref{tab:resutls_table} summarizes the results obtained by configuration generation and evaluation by functional correctness by exact match with human generated terraform configuration for two models.
\begin{enumerate}
% \item \textbf{GPT3.5-turbo:} We used in-context learning for the model to generate the configuration for all 49 tasks in AWS provider. 1-sample and 50-sample configurations were generated for each of the 49 tasks using two temperatures 0.2 and 0.6. These configurations were evaluated with human generated configurations by an functional correctness criterion. As can seen from the results, 59.16\% accuracy was obtained with 1-sample and 56.81\% accuracy was obtained by aggregating over 50-samples.

\item \textbf{GPT3.5-turbo:} Model used in-context learning to generate 1-sample and 50-sample configurations for all 49 tasks in AWS provider at temperature setting 0.2. The generated configurations are evaluated against human generated configurations using functional correctness by exact match. It obtained an average success rate(accuracy) of 59.16\%  with 1-sample and 56.81\%  with 50-samples, by calculating the mean of success rate of total number of samples generated under each task and the final score is the average of total success rate of all 49 tasks in AWS provider.

\item \textbf{Codeparrot:} In this also, model generated 1-sample and 50-sample configurations for all 49 tasks in AWS provider and evaluated the generated configurations with human generated configurations using functional correctness by exact match. The same methodology mentioned above in GPT3.5-turbo for calculating the average success rate(accuracy) is followed and obtained an accuracy of 8.2\% with 1-sample and 8\% accuracy with 50-samples.

% In this, the model generated 1-sample for each of the tasks and compared all 49 tasks with the human Terraform configurations with the exact match procedure and measured the average success rate. Similarly, the same approach was done with 50-samples for each of the 49 tasks and the average success rate was calculated. As can seen from the results table, 8.2\% accuracy was obtained with 1-sample and 8\% accuracy was obtained by aggregating over 50-samples.
\end{enumerate}

In summary, GPT-3.5-Turbo outperforms the CodeParrot model due to its extensive and diverse training dataset, as well as its inherent adaptability through fine-tuning that collectively contributed to GPT-3.5 Turbo's remarkable performance.

% In summary, GPT-3.5 Turbo outperforms the CodeParrot model. This superior performance can be attributed to several factors, including its extensive and diverse training dataset, as well as its inherent adaptability through fine-tuning. These aspects collectively contribute to GPT-3.5 Turbo's remarkable performance compared to CodeParrot.

\begin{table}[t]
\centering\small
\begin{tabular}{ccc}
\toprule
                                                                   & \textbf{Single sample} & \textbf{50 samples} \\ \midrule
GPT-3.5 turbo                                                      & 59.18\%                                                  & 56.81\%    \\ \midrule
\begin{tabular}[c]{@{}l@{}}CodeParrot \\ (Small 110M)\end{tabular} & 8.2\%                                                    & 8\%        \\ \bottomrule
\end{tabular}
\caption{Experimental Results}
\label{tab:resutls_table}
\end{table}

\section{IaC using LLM: Safety and Ethical Considerations}
This section outlines the safety and ethical considerations of LLM-generated configurations in production environments while proposing potential resolutions.

% Infrastructure as Code generated by LLMs have some challenges wrt the accuracy of the configurations generated and safety of using these in production environments. This section outlines the safety and ethical considerations for generation of IAC using LLMs and possible resolutions

% \subsection{Safety Considerations and Best Practices}

\subsection{\textbf{Safety Concerns:}}
\begin{itemize}
\item \textbf{Security Risks:} Unvalidated IaC can lead to vulnerabilities like misconfigured databases, lax security, and exposed secrets.

\item \textbf{Over-Reliance:} Blindly relying on LLM generated configurations is risky, understanding them is crucial.

\item \textbf{Resource Overutilization:} Poorly tuned IaC can create unnecessary resources, that results in cost overruns and also comes with the risk of over-provisioning leading to environmental expenses. 

\item \textbf{Updates and Maintenance:} LM's struggle with real-time changes to cloud platform etc., leading to outdated or ineffective setups.
\end{itemize}

% \begin{itemize}
% % \item \textbf{Security Risks:} The resulting IaC could be vulnerable if it is not properly validated. For instance, it might result in improperly setup databases, excessively permissive security groups, or exposed secrets.
% % \item \textbf{Over-Reliance:} It can be dangerous to rely only on LLM-generated configurations without comprehending them. Understanding what is being deployed is crucial.
% % \item \textbf{Resource Overutilization:} If the IaC generated is not properly tuned, it may produce resources that are larger or more numerous than required, which could result in cost overruns.
% % \item \textbf{Updates and Maintenance:} Unless they have just been trained, language models cannot keep up with real-time modifications to cloud platforms or software stacks. Deploying out-of-date or ineffective setups can result from following obsolete recommendations.
% % \end{itemize}
\subsection{\textbf{Best Practices:}}
\begin{itemize}
\item \textbf{Review and Validate:} Continuous evaluation of IaC for performance, security, and compliance through manual and automated reviews.

\item \textbf{Test in Isolated Environments:} Pre-deployment testing in sandbox environments helps uncover issues overlooked during code review.

\item \textbf{Version Control:} Store IaC in version-controlled repositories for collaboration, auditing, and easy change reversals.

\item \textbf{Educate the Team:} Ensure your team understands IaC principles and leverages LLMs as tools, to enhance expertise by staying updated with the platform and technology.

% \item \textbf{Stay Updated:} Keep updated with IaC platform and technology to recognize when LLM councel may be outdated.

\item \textbf{Limit Permissions:} Safeguard production deployments by restricting LLM access and having human supervision in your CI/CD or automation framework.

\item \textbf{Feedback Loops:} Create feedback mechanisms to refine LLM training and prompts based on IaC deployment results.
\end{itemize}

\subsection{Ethical Considerations:}
\begin{itemize}
\item \textbf{Transparency:} LLMs that conceal their decision-making process can cause due diligence concerns when using IaC models, as stakeholders often seek more details to understand the decisions.
% LLMs that act as "black box" hiding their decision-making process can raise due diligence concerns while using the models for IaC. Stakeholders often require reasons more than "model recommendations" to understand those infrastructure decisions.

% \begin{enumerate}
%     \item \textbf{Opaque Algorithms:} LLMs act as ``black boxes" hiding their decision-making process, potentially raising concerns about due diligence while using the models for IaC.
    
%     \item \textbf{Explanation:} Stakeholders may seek reasons behind infrastructure decisions, and simply citing "model recommendations" may not meet their need for transparency.
% \end{enumerate}

% \begin{enumerate}
%     \item \textbf{Opaque Algorithms:} LLMs function as ``black boxes", which makes it difficult to comprehend the logic behind their outputs. Without a clear understanding, using these models for IaC could be interpreted as a failure to exercise due diligence.
%     \item \textbf{Explanation:} It's possible that stakeholders will want to know why particular infrastructure decisions were made, and just stating ``because the model said so" may not be sufficient.
% \end{enumerate}

\item \textbf{Accountability:} Clear responsibility lines are essential for preventing large-scale failures or breaches in LLM-generated IaC, as determining accountability for defective and unsecure systems is complex.
% \begin{enumerate}
% \item \textbf{Who's Responsible?:} Determining accountability for defective and unsecure LLM-generated IaC is complex. Clarity on responsibility is crucial.

% \item \textbf{Error Propagation:} IaC errors can lead to large-scale failures or breaches. Clear responsibility lines are vital to avoid cascading problems.
% \end{enumerate}

% \begin{enumerate}
% \item \textbf{Who's Responsible?:} Who is accountable if an LLM generates defective or unsecure IaC? who controls the model? The architect who followed its suggestions? the business?
% \item \textbf{Error Propagation:} IaC oversights or mistakes can have a domino effect that could result in widespread failures or breaches. It is essential to ensure that responsibility lines are clear.
% \end{enumerate}

\item \textbf{Bias and Fairness:} LLMs trained on suboptimal data may perpetuate errors and overlook organizational or cultural nuances, creating IaC suitable for one context but not another.

\item \textbf{Dependency and Vendor Lock-In:} Excessive reliance on a single LLM for IaC risks vendor lock-in, reduced flexibility, and potential cost escalation.

% By significantly relying on a single LLM for IaC, enterprises risk becoming reliant on one vendor or model, losing flexibility, and sometimes even incurring higher costs.
\end{itemize}

\subsection{\textbf{Recommendations:}}
\begin{itemize}
\item \textbf{Human-in-the-loop:} Incorporating human judgment for critical infrastructure decisions.

\item \textbf{Data Diversity:} Ensuring varied training data for comprehensive best practices.

\item \textbf{Regular Audits:} Periodically reviewing LLM-generated IaC for bias and inefficiencies.

\item \textbf{Stakeholder Education:} Ensuring stakeholders understand LLM capabilities and limitations to manage expectations effectively.
\end{itemize}

% \begin{itemize}
% \item \textbf{Human-in-the-loop:} Employing human judgment in all decision-making, especially for critical infrastructure choices.

% \item \textbf{Diversity in Training Data:} Ensuring that LLM training data is diverse and covers a wide range of best practices.

% \item \textbf{Regular Audits:} Regularly examining LLM-generated IaC to identify and correct biases or inefficiencies.

% \item \textbf{Stakeholder Education:} Ensuring stakeholders understand LLM capabilities and limitations to manage expectations effectively.
% \end{itemize}

% \begin{itemize}
% \item \textbf{Human-in-the-loop:} Always using human judgment when making decisions, especially when it comes to important infrastructural choices.
% \item \textbf{Diversity in Training Data:} Making sure that the training data for LLMs is varied and encompasses a wide range of best practices.
% \item \textbf{Regular Audits:} Periodically reviewing and auditing the IaC produced by LLMs to spot biases or inefficiencies and fixing them.
% \item \textbf{Stakeholder Education:} Informing stakeholders on the capabilities and constraints of LLMs will help to maintain reasonable expectations.
% \end{itemize}
Based on our experimental study and detailed analysis from various references, we conclude that LLMs have the ability to generate IaC with great efficiency, but they must be employed carefully with awareness of their ethical ramifications.

\section{Challenges and Future Study}
Limited GitHub training data and Terraform representation may produce syntactically correct but erroneous code. LLMs lack awareness of current practices and security, risking data exposure. They may offer unsuitable solutions, misaligned with use cases. API updates affect code quality. Testing complex LLM-generated IaC complicates deployment. Lack of best practices and comments hampers modifications. Integration issues with DevOps tools can lead to cost inefficiencies.

To mitigate challenges, we can implement a comprehensive review process by engaging domain experts, and rigorously testing generated infrastructure code. In future, we can also use LLMs as assistants in multi-turn IaC Chatbots with automatic validations. Also we plan to expand our experiments using 1000 samples per task, comparing our results with open and closed-source models.

% These instructions are for authors submitting papers to *ACL conferences using \LaTeX. They are not self-contained. All authors must follow the general instructions for *ACL proceedings,\footnote{\url{http://acl-org.github.io/ACLPUB/formatting.html}} and this document contains additional instructions for the \LaTeX{} style files.

% The templates include the \LaTeX{} source of this document (\texttt{acl.tex}),
% the \LaTeX{} style file used to format it (\texttt{acl.sty}),
% an ACL bibliography style (\texttt{acl\_natbib.bst}),
% an example bibliography (\texttt{custom.bib}),
% and the bibliography for the ACL Anthology (\texttt{anthology.bib}).
% Entries for the entire Anthology, followed by custom entries
\bibliography{anthology, custom}
% \appendix

% \section{Example Appendix}
% \label{sec:appendix}

% This is an appendix.

\end{document}